\documentclass[12pt]{iopart}

\expandafter\let\csname equation*\endcsname\relax
\expandafter\let\csname endequation*\endcsname\relax
\usepackage{amsmath,amsfonts,amssymb}
\usepackage{bm}
\usepackage{iopams}  
\usepackage{cite}
\usepackage{url}
\usepackage{graphicx}
\usepackage{xcolor}
\usepackage{hyperref}
\usepackage{soul}

\newcommand{\kapt}{\tilde{\kappa}}
\newcommand{\mut}{\tilde{\mu}}
\newcommand{\rd}{{\rm d}}
\newcommand{\hu}{\hat{\bm u}}
\newcommand{\Pe}{{\rm Pe}}

\newcommand{\calX}{{\cal X}}
\newcommand{\calB}{{\cal B}}

\begin{document}

\title{Nonequilibrium energetics of sensing and actuation by 
a smart active particle}

\author{Luca Cocconi\textsuperscript{1,2}, Beno{\^i}t Mahault\textsuperscript{2} and Lorenzo Piro\textsuperscript{3}}
\address{
$^1$ Cavendish Laboratory, University of Cambridge, Cambridge CB3 0US, UK\\
$^2$ Max Planck Institute for Dynamics and Self-Organization (MPIDS), 37077 G{\"o}ttingen, Germany\\
$^3$ Department of Physics \& INFN, University of Rome ``Tor Vergata", Via della Ricerca Scientifica 1, 00133 Rome, Italy\\}

\ead{lc731@cam.ac.uk}
\vspace{10pt}
\begin{indented}
\item[]\today
\end{indented}

\begin{abstract}
    Smart active agents must allocate finite energetic resources across distinct functions, yet the underlying thermodynamic trade-offs remain poorly understood. Here, we introduce a minimal model of a self-steering particle with an internal polarity-cue sensor coupled to an external environmental field, decomposing its steady-state entropy production rate into locomotion, actuation, and sensing costs.
    This separation exposes an energetic bookkeeping structure underlying even the simplest form of embodied navigation.
    The emergence of Pareto fronts linking energetic expenditure to localisation precision and path-following performance shows that feedback-controlled active motion is constrained by quantitative thermodynamic bounds that persist across distinct task geometries.
\end{abstract}

%
%
%
\section{Introduction}

Many of the most promising applications of active matter, from targeted drug delivery by functionalised colloids \cite{liu2023colloidal,li2017micro} to swarm nanorobotics \cite{sun2023mean,dorigo2020reflections}, rely on the successful integration of self-actuation with the distributed recording, processing, and utilisation of information at the level of individual agents \cite{levine2023physics,goldman2024robot,stark2021artificial}. Biology offers a wealth of paradigmatic examples in which evolution has woven these functionalities together into remarkably robust behaviours. Classic cases include bacterial chemotaxis \cite{keller1971model,endres2008accuracy} and quorum sensing \cite{daniels2004quorum,PapenfortQS}, where motility and signalling combine either to rectify otherwise unbiased random motion along chemical gradients or to trigger changes in gene expression once a critical population density is exceeded.

Embedding similarly sophisticated information-processing capabilities into synthetic active agents presents a formidable engineering challenge \cite{liu2023colloidal}. This difficulty arises not only from the demands of miniaturisation \cite{ju2025technology,chen2025roadmap,miskin2020electronically}, but also from the increasingly stringent physical constraints encountered as the characteristic energy scales of actuation, sensing, and thermal fluctuations approach one another \cite{bo2015thermodynamic,lan2012energy,sartori2015thermodynamics,yu2022energy} (whereas at larger scales the former typically dominates). Under the additional and natural assumption of a limited energetic budget \cite{yang2021physical}, it is therefore compelling to ask what generic trade-offs underlie such complex behaviours and, having identified them, how finite thermodynamic resources should be allocated to optimise specific performance metrics \cite{cocconi2025dissipation,welker2026accuracy,olsen2026information}.

Addressing these questions requires a consistent nonequilibrium energetic framework in which all relevant features of the physical model are treated on equal footing. In this respect, \textit{ad hoc} multilevel descriptions, such as those invoking autonomous control via machine-learning modules external to the dynamics \cite{muinos2021reinforcement,putzke2023optimal}, may be dynamically informative, but are often thermodynamically ill-posed. The development of analytically tractable models that coherently integrate autonomous actuation, sensing, and decision-making is thus a crucial step towards elucidating the physics of ``smart" active matter.

To advance this programme, we introduce a minimal model of a self-steering motile agent equipped with an internal polarity-cue sensor coupled to an external environmental field, such as a concentration or light-intensity gradient (Sec.~\ref{sec:model_setup}). 
Using tools from stochastic thermodynamics \cite{seifert2012stochastic}, we decompose the mean rate of energy dissipation into distinct contributions associated with self-propulsion, steering, and sensing. Having determined how these components partition the overall energetic budget across different parameter regimes (Sec.~\ref{s:epr_framework}),
we comment on how the contribution associated with sensing can be related to the rate at which information is collected by the sensor itself (Sec.~\ref{s:info_flow}).
As concrete illustrations, we examine the resulting dissipation–accuracy trade-offs in three prototypical tasks: localisation at a point target, localisation within a target disc, and navigation along a prescribed path (Sec.~\ref{s:example_policies}).

\section{Self-steering active particles with explicit sensory feedback }\label{sec:model_setup}

\begin{figure}[t!]
    \centering
    \includegraphics[width=0.7\linewidth]{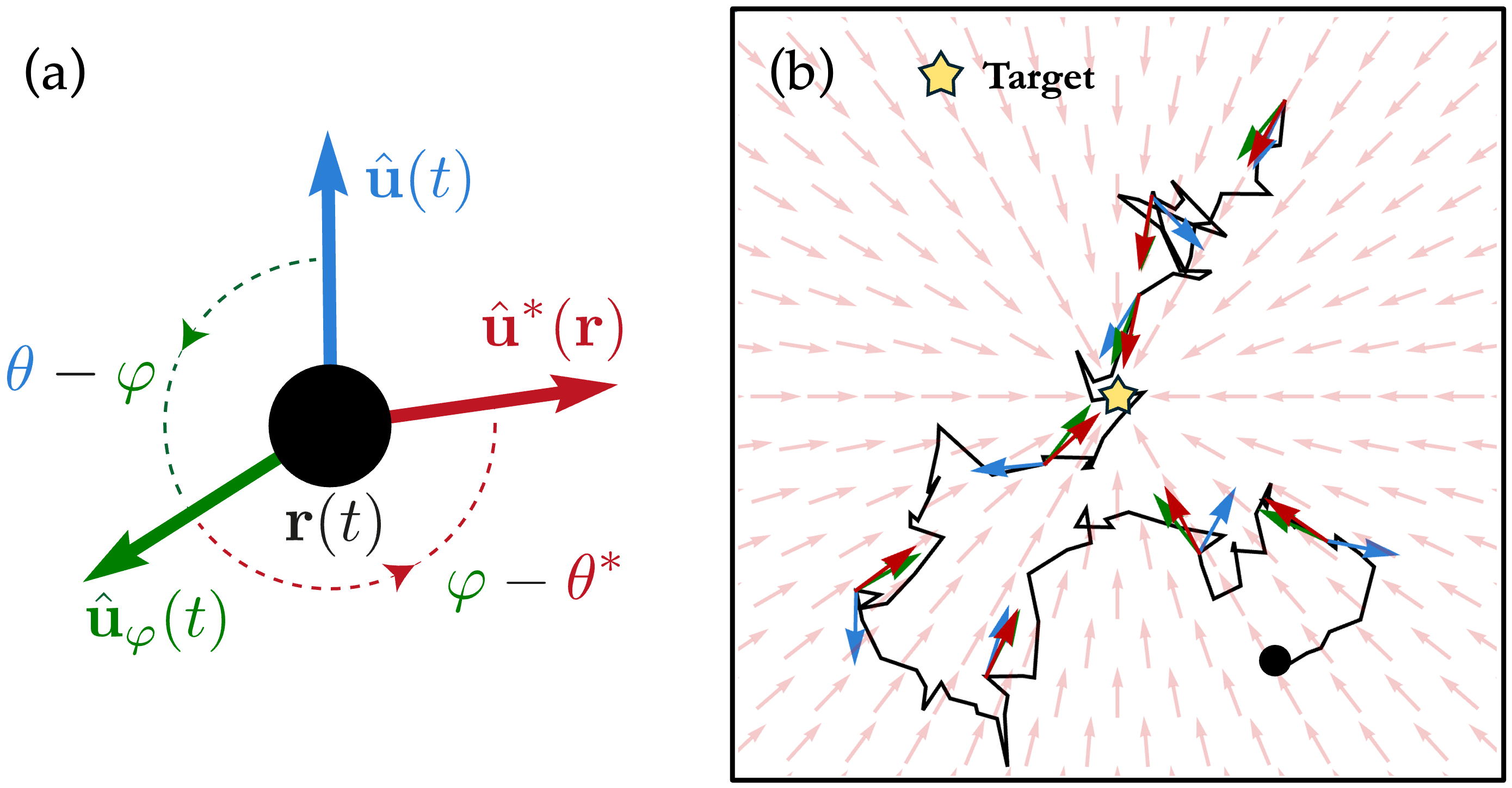}
    \caption{(a) Schematic illustration of the model of a smart active Brownian particle with explicit sensing and actuation. The particle self-propels at constant speed along its instantaneous heading direction $\hat {\bm u}(t)$ (blue vector). An internal polarity sensor with readout $\hat {\bm u}_\varphi(t)$ (green vector) aligns stochastically with a position-dependent steering policy $\hat {\bm u}^*({\bm r})$ (red vector), encoding information about the external environment. The heading direction is in turn actively steered toward the sensor readout, closing a feedback loop between sensing and actuation. (b) Exemplary particle trajectory (black) in the point-target setup, where the steering policy $\hat {\bm u}^*(r)$ (red vector field in the background) points toward a fixed target position denoted by the yellow star. At selected points along the trajectory, the three unit vectors shown in panel (a) --- the heading direction $\hat {\bm u}(t)$, the sensor readout $\hat {\bm u}_\varphi(t)$, and the steering policy $\hat {\bm u}^*({\bm r})$ --- are displayed. The trajectory was obtained from numerical integration of the Langevin equations with parameters $\tilde{\mu}=50$, $\tilde{\kappa}=1$, $\sigma_\varphi^2=0.01$, and $\Pe=10$.}
    \label{fig:schematic}
\end{figure}

We consider an agent moving in two dimensions at constant self-propulsion speed $w$.
Its state is specified by the center-of-mass position $\bm{r}(t)$ and by a heading direction $\hat{\bm{u}}_{\theta}(t) = (\cos\theta(t),\sin\theta(t))$, with $\theta \in [0,2\pi)$. Circumflexes will henceforth indicate unit vectors.
The agent is additionally equipped with an internal state $\hat{\bm{u}}_\varphi(t)$, corresponding to the readout of a polarity cue sensor. This sensor is coupled to a prescribed, stationary vector field $\hat{\bm{u}}^*(\bm{r}) = (\cos\theta^*(\bm{r}),\sin\theta^*(\bm{r}))$, which encodes the locally preferred direction of motion in space and we thus refer to it as the \emph{steering policy} (see Fig.~\ref{fig:schematic} for a schematic illustration).
The dynamics of the position, heading angle, and sensor readout are governed by the following set of overdamped Langevin equations:
\begin{subequations}
\label{eq:governing_langevin}
\begin{align}
    \dot{\bm{r}}(t) &= w \hat{\bm{u}}_\theta(t) + \sqrt{2D_r} \bm{\eta}_r(t)\\
    \label{eq:governing_langevin_theta}
    \dot{\theta}(t) &= \kappa \Gamma_\theta(\theta,\varphi)+ \sqrt{2D_\theta} \eta_\theta(t) \\
    \dot{\varphi}(t) &= \mu \Gamma_\varphi(\varphi,\theta^*(\bm{r})) + \sqrt{2D_\varphi} \eta_\varphi(t) \, ,
\end{align}
\end{subequations}
where the $\eta_{i}$ ($i\in \{x,y,\theta,\varphi\}$) are independent and identically distributed Gaussian white noises of unit covariance, while $D_r$, $D_\theta$, and $D_\varphi$ denote the one translational and two angular diffusion coefficients, respectively.

The drift term $\kappa \Gamma_\theta$ represents an internally generated torque that steers the heading angle $\theta$ toward the sensor readout $\varphi$, thereby quantifying the strength of actuation. 
Conversely, the term $\mu \Gamma_\varphi$ drives the sensor variable toward alignment with the local policy direction $\theta^*(\bm r)$, with $\mu$ setting the sensing rate. 
The microscopic origin of this alignment may be equilibrium or nonequilibrium in nature; here, it is treated at the level of effective dynamics.
For concreteness, we assume dipolar alignment interactions and set $\Gamma_\theta(s',s'') = \Gamma_\varphi(s',s'')= - \sin(s'-s'')$.
In the absence of steering ($\kappa = 0$), the positional dynamics of our model reduces to those of an active Brownian particle (ABP) in two dimensions, a paradigmatic model of active matter \cite{solon2015active}. Related models without an explicit sensing degree of freedom were analysed in Refs.~\cite{cocconi2025dissipation} and \cite{goh2022noisy}.

Activity enters our model through three distinct, independent mechanisms. First, persistent self-propulsion breaks the equilibrium fluctuation-dissipation relation \cite{de2013non}, representing the canonical form of activity in active matter. This contribution vanishes in the limit $w = 0$. 
Second, the nonreciprocal coupling between heading angle $\theta$ and sensor readout $\varphi$---where the heading ``chases'' the sensor but not vice versa---violates microscopic reversibility, with reciprocity restored only in the decoupled limit $\kappa=0$. Third, and more subtly \cite{chen2025numerical}, nonequilibrium persists even for $w=\kappa=0$: the unidirectional link from position $\bm{r}$ to sensor dynamics via $\Gamma_\varphi(\varphi,\theta^*(\bm{r}))$ couples spatial exploration to internal sensory reorientation, precluding detailed balance.

In the following, we rescale space and time as $\bm{r} \to (w/D_\theta)\bm{r}$ and $t \to t/D_\theta$, and introduce the dimensionless P{\'e}clet number ${\rm Pe} \equiv w^2/(D_r D_\theta)$, alignment strengths $\kapt \equiv \kappa/D_\theta$ and $\mut \equiv \mu/D_\theta$, and sensor accuracy $\sigma^2_\varphi = D_\varphi/(D_\theta\mut)$. Assuming that a statistical steady state for the process exists, the stationary probability density $P(\bm{r},\theta,\varphi)$ satisfies the Fokker-Planck equation
\begin{equation}\label{eq:fp_stationary}
    0 = \bm{\nabla} \cdot \bm{J}_r + \partial_\theta J_\theta + \partial_\varphi J_\varphi
\end{equation}
with the (generically non-vanishing) probability currents 
\begin{subequations}
\begin{align}
    \bm{J}_r(\bm{r},\theta,\varphi) &= [\hat{\bm{u}}_\theta - \Pe^{-1}\bm{\nabla}]P(\bm{r},\theta,\varphi) \\
    J_\theta(\bm{r},\theta,\varphi) &= [-\kapt\sin(\theta-\varphi)-\partial_\theta]P(\bm{r},\theta,\varphi) \label{eq:curr_theta}\\
    J_\varphi(\bm{r},\theta,\varphi) &= \mut[-\sin(\varphi-\theta^*(\bm{r}))-\sigma^2_\varphi\partial_\varphi]P(\bm{r},\theta,\varphi)~. \label{eq:curr_varphi}
\end{align}
\end{subequations}

\section{Thermodynamic costs of locomotion, actuation, and sensing}\label{s:epr_framework}

We equip the dynamics with a thermodynamic interpretation by computing the mean rate of entropy production at steady state \cite{cocconi2020entropy,seifert2012stochastic}. This is proportional to the mean stochastic work done by the generalized active forces, and can be written in our rescaled units as
\begin{equation}\label{eq:epr_full}
    \dot{\Sigma} = \int \rd\bm{r} \rd\theta \rd\varphi \ \left[\Pe \bm{J}_r \cdot \hat{\bm{u}}_\theta - \kapt \sin(\theta-\varphi)J_\theta - \frac{1}{\sigma_\varphi^2} \sin(\varphi-\theta^*(\bm{r}))J_\varphi \right]~.
\end{equation}
The first term in the integrand represents the dissipation from self-propulsion and evaluates exactly to $\dot{\Sigma}_{\rm prop}=\Pe$, independent of the steering policy or model parameters. The remaining terms, however, depend non-trivially on the stationary density $P(\bm{r},\theta,\varphi;\hat{\bm{u}}^*)$; we derive their approximate forms below under specific conditions.

For clarity, we introduce the shorthand
\begin{subequations}
\label{eq:epr_shorthand_contribs}
\begin{align}
    \dot\Sigma_{\rm act} &= -\kapt\int \rd\bm{r} \rd\theta \rd\varphi \  \sin(\theta-\varphi)J_\theta~, \\
    \dot\Sigma_{\rm sens} &= - \frac{1}{\sigma_\varphi^2}\int \rd\bm{r} \rd\theta \rd\varphi \  \sin(\varphi-\theta^*(\bm{r}))J_\varphi~, \label{eq:sigma_sens_firstform}
\end{align}
\end{subequations}
where $\dot\Sigma_{\rm act}$ captures the thermodynamic cost of actuation/feedback control, while $\dot\Sigma_{\rm sens}$ quantifies the cost of sensing. Before proceeding, we note that by multiplying both sides of Eq.~\eqref{eq:fp_stationary} by $\cos(\varphi-\theta^*)$ and integrating over all degrees of freedom, one obtains the following alternative form for $\dot\Sigma_{\rm sens}$:
\begin{equation}\label{eq:epr_sens_alt}
    \dot\Sigma_{\rm sens} = - \frac{1}{\sigma_\varphi^2} \left[ \langle \hat{\bm{u}}_\theta \cdot \nabla \cos(\varphi-\theta^*(\bm{r}))\rangle + {\rm Pe}^{-1} \langle \nabla^2 \cos(\varphi-\theta^*(\bm{r}))\rangle \right]~,
\end{equation}
where angular brackets $\langle \cdot \rangle$ henceforth denote averages with respect to $P(\bm{r},\theta,\varphi;\hat{\bm{u}}^*)$.

\subsection{Fast-sensing limit and effective dynamics}
\label{sec:approximations}

To derive analytical expressions for the stationary density $P(\bm{r},\theta,\varphi;\hat{\bm{u}}^*)$---and thus explicit formulas for the entropy production contributions $\dot\Sigma_{\rm act}$ and $\dot\Sigma_{\rm sens}$---we establish a hierarchy of timescales and apply a multiscale approach \cite{pavliotis2008multiscale} as follows. 
In particular, we consider the regime where sensing happens on fast timescales,
which amounts to taking the limit
$\mut \to \infty$ while keeping the sensor precision $\sigma_\varphi^2$ and all other parameters constant. As a result, we have from the Fokker-Planck equation \eqref{eq:fp_stationary} that, to leading order in small $\mut^{-1}$, the joint probability factorises
\begin{equation}\label{eq:p_factorised}
    P(\bm{r},\theta,\varphi;\hat{\bm{u}}^*) =  P_{\rm eff}(\bm{r},\theta;\hat{\bm{u}}^*) P_\varphi(\varphi;\hat{\bm{u}}^*) + \mathcal{O}(\mut^{-1})
\end{equation}
with 
\begin{equation}\label{eq:marginal_varphi}
    P_\varphi(\varphi;\hat{\bm{u}}^*)= \frac{1}{2\pi I_0(\sigma_\varphi^{-2})} e^{\cos(\varphi-\theta^*(\bm{r}))/\sigma_\varphi^2}
\end{equation}
the Boltzmann weight obtained by solving $J_\varphi=0$ at constant $\bm{r}$. Here, $I_n$ is the modified Bessel function of the first kind of order $n$. 
This marginal also holds, independently of the fast-sensing assumption, for ``pinned'' problems (e.g., with external potentials halting motion) or spatially uniform policies $\hat{\bm{u}}^*$. From Eq.~\eqref{eq:p_factorised}, the key trigonometric moments relative to the policy read
\begin{subequations}
\label{eq:trig_moms_varphi}
\begin{align}
    \langle\cos(\varphi-\theta^*)\rangle &= c_0(\sigma_\varphi^{-2}) \\
    \langle\sin(\varphi-\theta^*)\rangle &= 0 \\
    \langle\cos^2(\varphi-\theta^*)\rangle &= 1 - \langle\sin^2(\varphi-\theta^*)\rangle = 1-\sigma_\varphi^{2}c_0(\sigma_\varphi^{-2}) 
\end{align}
\end{subequations}
where we defined $c_0(s)=I_1(s)/I_0(s)$ with $0\leq c_0(s)<1$. In the limit of fast sensing, the effective dynamics of the heading direction may be written as
\begin{align}
    \dot{\theta}(t) &= \sqrt{2}\eta_\theta(t) - \int \rd\varphi P_\varphi(\varphi;\hat{\bm{u}}^*) \tilde{\kappa} \sin(\theta-\varphi) \nonumber \\
    &= - \tilde{\kappa}c_0(\sigma_\varphi^{-2}) \sin(\theta-\theta^*(\bm{r})) + \sqrt{2}\eta_\theta(t)~,
    \label{eq:langevin_theta_fast}
\end{align}
mirroring the original Eq.~\eqref{eq:governing_langevin_theta} but with renormalised alignment $\kapt \to \kapt_{\rm eff} \equiv \kapt c_0(\sigma_\varphi^{-2})$, with $\kapt_{\rm eff}\leq \kapt$, and direct coupling from policy $\theta^*(\bm{r})$ to heading $\theta$.

Having established the relaxation of the probability density of the sensor readout as the fastest timescale, we now proceed to introduce an approximation scheme based on a gradient expansion of $P_{\rm eff}(\bm r,\theta)$ in Eq.\eqref{eq:p_factorised}. 
This enables closed-form evaluation of the steady-state moments in Eq.~\eqref{eq:epr_full} whenever spatial variations in $P_{\rm eff}$ remain smooth, effectively imposing a second adiabatic approximation: angular equilibration in $\theta$ outpaces positional dynamics in $\bm{r}$. To leading order in $\mut^{-1}$ and in gradients, the marginal positional density $\rho(\bm{r})\equiv\int\rd\theta P_{\rm eff}(\bm{r},\theta)$ obeys the effective stationary Fokker-Planck equation
\begin{equation}\label{eq:eff_fp_rho}
     \nabla\cdot \left[\left( \langle\bm{u}_\theta\rangle_\theta - {\rm Pe}^{-1}\bm{\nabla}\right)\rho({\bm r})\right] = 0
\end{equation}
with the position-dependent average heading direction
\begin{align}
\langle \hu_\theta \rangle_\theta & = c_0(\kapt_{\rm eff})\hu^* - \left[ c_1(\kapt_{\rm eff})\bm I + c_2(\kapt_{\rm eff})\hu^*\hu^* \right]\cdot\nabla\ln\rho  \nonumber\\
& + c_3(\kapt_{\rm eff})\hu^*(\nabla\cdot\hu^*) + c_4(\kapt_{\rm eff})\hu^*_\perp(\nabla\cdot\hu^*_\perp) + {\cal O}(\nabla^2,\mut^{-1})~, \label{eq:ave_heading_momexp}
\end{align}
where $\bm I$ is the identity matrix, while $\mathcal{O}(\cdot)$ encompasses gradients of the density, policy, or flow, and the coefficients $c_{1,2,3,4}$ are detailed in \ref{app:mom_exp}. 
Inserting \eqref{eq:ave_heading_momexp} into \eqref{eq:eff_fp_rho} shows that the combination of sensing and actuation induces: (i) an effective drift along the local policy $\hu^*(\bm{r})$ with a renormalised speed $c_0(\kapt_{\rm eff})<1$ in units of the bare self-propulsion speed; (ii) an anisotropic renormalisation of the diffusion tensor, with enhanced fluctuations along the axis defined by $\theta^*(\bm{r})$ ($c_2(\kapt_{\rm eff}) > 0$).
From Eq.~\eqref{eq:ave_heading_momexp}, we additionally note that
\begin{equation}\label{eq:cos_theta_mon}
    \langle \cos(\theta-\theta^*(\bm{r}))\rangle = \langle \hu_\theta \cdot \hu^* \rangle = c_0(\kapt_{\rm eff}) + c_3(\kapt_{\rm eff}) \langle \nabla\cdot\hu^*\rangle + O(\nabla^2,\mut^{-1})~.
\end{equation}
Lastly, the moment expansion yields the approximate identity
\begin{equation}\label{eq:cos_sin2_id}
     \kapt_{\rm eff}^2\langle\sin^2(\theta-\theta^*(\bm{r})) \rangle - \kapt_{\rm eff}\langle\cos(\theta-\theta^*(\bm{r})) \rangle = - c_0(\kapt_{\rm eff})\langle \nabla \cdot \hu^*\rangle + O(\nabla^2,\mut^{-1})~,
\end{equation}
derived in \ref{app:mom_exp}.

\subsection{Approximate analytical forms for actuation and sensing costs} 
\label{sec:epr_compact}

Combining Eqs.~\eqref{eq:curr_theta} and~\eqref{eq:epr_shorthand_contribs}, by adding and subtracting $\theta^*$ in the argument of the trigonometric functions, we can write the contribution to the entropy production associated with actuation as
\begin{align}
    \dot\Sigma_{\rm act} &=  \kapt^2 \langle \sin^2[(\theta-\theta^*)-(\varphi-\theta^*)] \rangle - \kapt \langle \cos[(\theta-\theta^*)-(\varphi-\theta^*)])\rangle  \nonumber \\
    &= \kapt^2 \sigma_\varphi^2 c_0(\sigma_\varphi^{-2}) + \kapt^2[1-2 \sigma_\varphi^2 c_0(\sigma_\varphi^{-2})]\langle \sin^2(\theta-\theta^*)\rangle - \kapt c_0(\sigma_\varphi^{-2}) \langle \cos(\theta-\theta^*)\rangle
\end{align}
where we have used Eq.~\eqref{eq:trig_moms_varphi} to evaluate the trigonometric moments with argument $(\varphi-\theta^*)$, which factorise based on Eq.~\eqref{eq:p_factorised}. Substituting Eqs.~\eqref{eq:cos_theta_mon} and \eqref{eq:cos_sin2_id} into this expression allows us to get rid of the remaining trigonometric moments, leading to
\begin{equation}\label{eq:epr_act}
    \dot\Sigma_{\rm act} = \dot\Sigma_{\rm act,pinned} + \dot\Sigma_{\rm act,\nabla} + \mathcal{O}(\nabla^2,\mut^{-1})
\end{equation}
where we have split this contribution into two parts 
\begin{subequations}
\label{eq:epr_act_explicit}
\begin{align}
    \dot\Sigma_{\rm act,pinned} &= \frac{\kapt_{\rm eff}^2 \sigma_\varphi^2}{c_0(\sigma_\varphi^{-2})} + \kapt_{\rm eff} \left( \frac{1-2\sigma_\varphi^2c_0(\sigma_\varphi^{-2})}{c_0^2(\sigma_\varphi^{-2})}-1\right)c_0(\kapt_{\rm eff}) \label{eq:epr_act_pinned}\\
    \dot\Sigma_{\rm act,\nabla} &= -c_0(\kapt_{\rm eff})
    \left[ 1 - \left( 1 - \frac{\kapt_{\rm eff} c_3(\kapt_{\rm eff})}{c_0(\kapt_{\rm eff})}\right)
    \left( \frac{1}{c_0^2(\sigma_\varphi^{-2})} - \frac{2\sigma_\varphi^2}{c_0(\sigma_\varphi^{-2})} - 1\right)\right] \langle \nabla \cdot \hu^*\rangle \label{eq:epr_act_nabla}
\end{align}
\end{subequations}
of order zero and one in $\nabla$, respectively. 
As the notation suggests, the contribution $\dot\Sigma_{\rm act,pinned}$ remains finite in the ``pinned'' problem ($\dot{\bm{r}}=0$), as well as when the policy is trivial, i.e., uniform $\theta^*({\bm r})={\rm const}$. The existence of such a contribution indicates that, in the limit of fast sensing, a constant thermodynamic cost is incurred by the actuation mechanism as a result of its continuous attempt to enforce alignment of the heading angle to a fluctuating sensor readout, irrespective of the details of the steering policy \cite{sartori2014thermodynamic}. 
On the other hand, the contribution $\dot\Sigma_{\rm act,\nabla} \propto - \langle \nabla \cdot \hu^*\rangle$, which vanishes for homogeneous policies, formalizes the intuition that policies designed to localise the swimmer in the proximity of a target subspace inevitably incur a thermodynamic cost since density accumulates in regions of negative policy divergence (in the parameter regime of interest, the sign of the policy-independent prefactor is positive). Note that evaluating \eqref{eq:epr_shorthand_contribs} only requires knowledge of the marginal density $\rho(\bm{r})$, rather than the full joint density $P({\bm r},\theta,\varphi)$, cf.\ Eq.~\eqref{eq:epr_shorthand_contribs}.

For sensing, the factorisation \eqref{eq:p_factorised}, combined with Eq.~\eqref{eq:epr_sens_alt} and \eqref{eq:trig_moms_varphi}, yields the exact (to all orders in $\nabla$) fast-sensing form
\begin{equation}
    \dot\Sigma_{\rm sens} =  \frac{c_0(\sigma_\varphi^{-2})}{\sigma_\varphi^2\Pe }  \langle |\nabla \theta^*(\bm{r})|^2\rangle + \mathcal{O}(\mut^{-1})\ . \label{eq:epr_sens_fast}
\end{equation}
Though formally higher order in gradients than our $\dot{\Sigma}_{\rm act}$ truncation, its convergence properties---explored below for localisation at a point target---reveal subtleties in the positional density $\rho$. Once again, the evaluation of \eqref{eq:epr_sens_fast} only requires knowledge of the marginal $\rho(\bm{r})$.

\section{Thermodynamic cost of sensing as an information flow}\label{s:info_flow}

Given that $\varphi$ represents an internal polarity sensor coupled to the steering policy $\hu^*$, we anticipate a link between its associated entropy production, $\dot\Sigma_{\rm sens}$, and the rate of information acquisition by the sensor. To quantify this energetics-informatics connection, we evaluate the steady-state multivariate information flow $\dot{I}_{\to \varphi}$ into $\varphi$ from the remaining degrees of freedom~\cite{allahverdyan2009thermodynamic,loos2020irreversibility}:
\begin{align}
    \dot{I}_{\to \varphi} 
    &= \int d\bm{r} d\theta d\varphi \ \ln \frac{P(\varphi)}{P(\bm{r},\theta,\varphi)} \partial_\varphi J_\varphi ~.
\end{align}
This is a signed quantity that vanishes at equilibrium and quantifies whether the rest of the system is gaining control over/sending information to ($\dot{I}_{\to \varphi}>0$) or losing control over/receiving information from ($\dot{I}_{\to \varphi}<0$) the variable $\varphi$.
Analogous expressions exist for the information flows into $\bm{r}$ and $\theta$. Together, these sum to the time derivative of the multivariate mutual information, $\sum_{s \in \{\bm{r},\theta,\varphi\}}\dot{I}_{\to s} = \dot{I}$ \cite{loos2020irreversibility}, defined here according to the convention \cite{srinivasa2005review}
\begin{align}
    I[P(\bm{r},\theta,\varphi)] &= \int d\bm{r} d\theta d\varphi \ P(\bm{r},\theta,\varphi) \ln \frac{P(\bm{r},\theta,\varphi)}{\rho(\bm{r})P(\theta)P(\varphi)} \nonumber \\
    &= {\rm KL}[P(\bm{r},\theta,\varphi)||\rho(\bm{r})P(\theta)P(\varphi)] \, ,
\end{align}
where KL denotes the Kullback-Leibler divergence \cite{kullback1951information}, taken between the joint distribution and the product of its marginals. 
In the fast sensing limit, using Eq.~\eqref{eq:p_factorised},
\begin{align}
\dot{I}_{\to \varphi} 
    &= \int d\bm{r} d\theta d\varphi \ \left[ \ln \frac{P(\varphi)}{P(\varphi;\hu^*)} + \ln \frac{1}{P_{\rm eff}(\bm{r},\theta;\hu^*)}\right]\partial_\varphi J_\varphi \nonumber \\
    &=\int d\bm{r} d\theta d\varphi \ \ln \frac{P(\varphi)}{P(\varphi;\hu^*)}[-\bm{\nabla}\cdot \bm{J}_r - \partial_\theta J_\theta]~, \label{eq:inf_flow_varphi}
\end{align}
where we have used Eq.~\eqref{eq:fp_stationary} and the fact that $P_{\rm eff}$ is independent of $\varphi$.

Upon further manipulation of Eq.~\eqref{eq:inf_flow_varphi}, we find
\begin{align}
    \dot{I}_{\to \varphi}  &= - \int d\bm{r} d\varphi \ \ln \frac{P(\varphi)}{P(\varphi;\hu^*)} \bm{\nabla}\cdot \left( \int d\theta \ \bm{J}_r \right) + \mathcal{O}(\mut^{-1}) \nonumber\\
    &= \frac{1}{\rm Pe} \int d\bm{r} d\varphi \ \ln \frac{P(\varphi)} {P(\varphi;\hu^*)}  \rho(\bm{r}) \nabla^2 P(\varphi;\hu^*) + \mathcal{O}(\mut^{-1}) \, ,
\end{align}
leveraging $\theta$-independence of $P(\varphi;\hu^*)$ and symmetries in its Boltzmann form~\eqref{eq:marginal_varphi}.
Also, through Eq.~\eqref{eq:marginal_varphi} we get
\begin{equation}
    \nabla^2 P(\varphi;\hu^*) = \frac{P(\varphi;\hu^*)}{\sigma_\varphi^2} \left[  \frac{\sin^2(\varphi-\theta^*)}{\sigma_\varphi^2} - \cos(\varphi-\theta^*)\right] |\nabla \theta^*(\bm{r})|^2 \, ,
\end{equation}
and so overall
\begin{equation}\label{eq:info_flow_final}
    \dot{I}_{\to \varphi}  = \frac{\mathcal{C}(\sigma_\varphi^2)}{\sigma_\varphi^2 {\rm Pe} } \left\langle |\nabla \theta^*(\bm{r})|^2  \right\rangle + \mathcal{O}(\mut^{-1}) \, ,
\end{equation}
where 
\begin{equation}\label{eq:mathaclC_def}
    \mathcal{C}(\sigma_\varphi^2)\equiv \int d\varphi \ P(\varphi;\hu^*)\ln \frac{P(\varphi)} {P(\varphi;\hu^*)} \times \left[  \frac{\sin^2(\varphi-\theta^*)}{\sigma_\varphi^2} - \cos(\varphi-\theta^*) \right] = c_0(\sigma_\varphi^{-2})
\end{equation}
is actually policy-independent (via a change of variable $\varphi \to \varphi'+\theta^*$) and thus factors from the average. 
Comparing with Eq.~\eqref{eq:epr_sens_fast}, we thus find that the cost of sensing and the stationary information flow into the sensor are exactly equal, $\dot\Sigma_{\rm sens}=\dot{I}_{\to \varphi}$, for all values of $\sigma_\varphi^2$ in the fast sensing limit.
The bound $\eta = \dot{I}_{\to \varphi}/\dot{\Sigma}_{\rm sens} \leq 1$ on the efficiency of information flow derived in Ref.~\cite{loos2020irreversibility} is thus saturated in this system.

\section{Unified thermodynamic constraints across navigation tasks} \label{s:example_policies}

\subsection{Localisation at a point target}
\label{sec:target_point}

We first apply our framework to the paradigmatic case of a stationary point target, defined by the steering policy $\hu^*(\bm{r})=-(\bm{r}-\bm{r}_{\rm target})/|\bm{r}-\bm{r}_{\rm target}|$ designed to localise the agent in the proximity of $\bm{r}_{\rm target}$. 
This serves as a minimal model for chemo- or phototaxis, with $\hu^*$ representing the normalised gradient of a chemoattractant concentration~\cite{keller1971model,endres2008accuracy} or light intensity~\cite{lozano2016phototaxis,frangipane2018dynamic}, respectively.
Translational invariance lets us set $\bm{r}_{\rm target}= \bm{0}$ without loss of generality, yielding $|\nabla\theta^*|=-\bm{\nabla}\cdot\hu^*=r^{-1}$ where $r \equiv |\bm{r}|$ denotes the radial displacement.

Substituting this policy into the effective Fokker-Planck equation~\eqref{eq:eff_fp_rho} and enforcing zero steady-state radial current gives
\begin{equation} \label{eq:pdf_rho_target_gen}
    \frac{\rd \ln \rho(\bm{r})}{\rd r}
    = - \frac{c_0(\kapt_{\rm eff}) + r^{-1} c_3(\kapt_{\rm eff})}{\Pe^{-1} + c_1(\kapt_{\rm eff})+c_2(\kapt_{\rm eff})}~.
\end{equation}
The radial displacement is thus Gamma-distributed 
\begin{equation}\label{eq:rho_target}
    \rho(r) \equiv \int \rd\bm{r} \delta(r-|\bm{r}|)\rho(\bm{r}) = \frac{r^{1+\eta} e^{-r/\ell}}{\ell^{2+\eta}\Gamma(2+\eta)}
\end{equation}
characterised by a cutoff length $\ell$ and an anomalous short-distance exponent $\eta$
\begin{equation}
    \ell = \frac{{\rm Pe}^{-1} + c_1(\kapt_{\rm eff}) + c_2(\kapt_{\rm eff})}{c_0(\kapt_{\rm eff})}, \quad\quad \eta = \frac{c_3(\kapt_{\rm eff})}{{\rm Pe}^{-1} + c_1(\kapt_{\rm eff}) + c_2(\kapt_{\rm eff})}~.
\end{equation}
The parameter $\ell$ sets the typical localization length, while $\eta$ controls the algebraic scaling of $\rho(r)$ near the origin. 
All steady-state observables entering the entropy production can be evaluated explicitly from the density profile~\eqref{eq:rho_target}. In particular,
\begin{align}\label{eq:theory}
    \langle |\nabla \theta^*|^2\rangle = \frac{\Gamma(\eta)}{\ell^2\Gamma(2+\eta)}, \quad \langle \nabla\cdot \hu^*\rangle = -\frac{\Gamma(1+\eta)}{\ell\Gamma(2+\eta)}, \quad \sigma_r^2 \equiv \langle r^2\rangle = \frac{\ell^2 \Gamma(4+\eta)}{\Gamma(2+\eta)}~.
\end{align}
Within the approximation scheme developed above, these expressions yield closed-form formulas for the actuation and sensing entropy production rates, Eqs.~\eqref{eq:epr_act_explicit} and~\eqref{eq:epr_sens_fast}. 

A notable feature emerges in the weak-alignment limit: while $\lim_{\kapt_{\rm eff}\to 0}\dot\Sigma_{\rm act}=0$, the sensing cost approaches a nontrivial constant:
\begin{equation}
    \lim_{\kapt_{\rm eff}\to 0} \dot\Sigma_{\rm sens} = \frac{16 c_0(\sigma_\varphi^{-2}) }{3 \sigma_\varphi^{2}  (2  + \Pe )}~,
\end{equation}
disappearing only in the simultaneous limit $\sigma_\varphi^2 \to \infty$ of vanishing sensor accuracy. Thus, even when steering is ineffective, maintaining a finite-precision sensory readout carries a persistent thermodynamic cost.

\begin{figure}[t!]
    \centering
    \includegraphics[width=\linewidth]{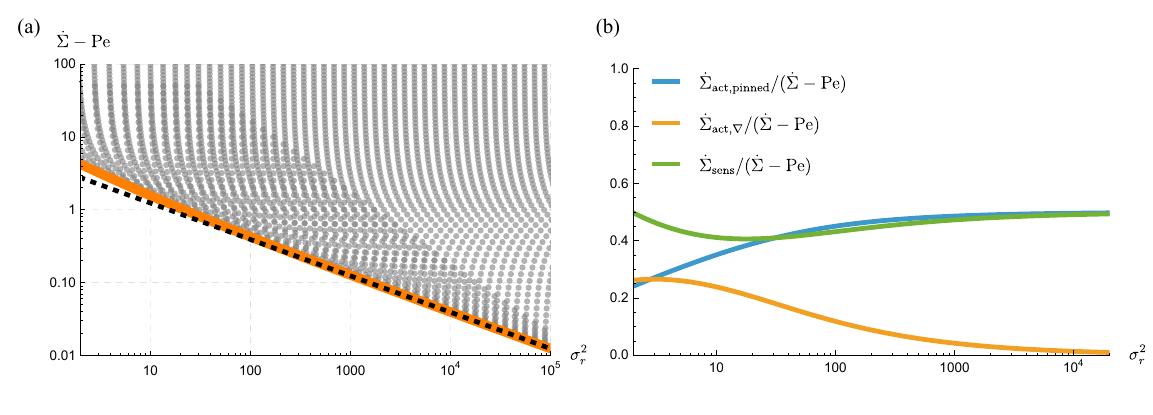}
    \caption{Dissipation-accuracy tradeoff for an active Brownian motion with polarity sensing and actuation tasked with localisation at a point target. (a) Numerical results for the entropy production and  variance of the radial displacement for different choices of $(\kapt,\sigma_\varphi^2)$ (grey scatter) are shown alongside the numerical (orange) and analytical (black dashed) Pareto fronts. Here, we set $\Pe=10$. (b) Relative contribution of the partial entropy production terms identified in Sec.~\ref{sec:epr_compact} to the total dissipation (excluding self-propulsion) along the Pareto front.
    }
    \label{fig:pareto_point_target}
\end{figure}

We can now compare these results with the mean squared displacement $\sigma_r^2$ to characterise the dissipation-accuracy tradeoffs occurring in this model. 
Since $\sigma_r^2$ depends on $\kapt$ and $\sigma_\varphi^2$ only via the effective alignment strength $\kapt_{\rm eff}$, we minimise the total dissipation $\dot\Sigma$ with respect to $\sigma_\varphi^2$ at fixed $\kapt_{\rm eff}$ to obtain an expression for the Pareto front associated with the multi-objective optimisation problem of balancing precision and energy expenditure. This optimisation admits an analytical solution for small $\kapt_{\rm eff}$, whereby we find $\sigma^2_{\varphi, \rm opt} = 1/\sqrt{3\kapt_{\rm eff}}$ and:
\begin{subequations}
\label{eq:small_kapt_tradeoff}
\begin{align}
    {\sigma_{r, \rm opt}^2}(\kapt_{\rm eff} ) &= \frac{6(2+{\rm Pe})^2}{{\rm Pe}^2 \kapt_{\rm eff}^2} + \mathcal{O}(1)\\
    \dot{\Sigma}_{\rm opt}(\kapt_{\rm eff}) &= \Pe + \frac{8 \kapt_{\rm eff}}{\sqrt{3(2+{\rm Pe})}}+ \mathcal{O}(\kapt_{\rm eff}^2)~.
\end{align}
\end{subequations}
At fixed $\Pe$, these expressions imply that the optimal dissipation and accuracy are related for weak alignment by $\dot{\Sigma}_{\rm opt} - \Pe \simeq \sigma_{r, \rm opt}^{-1}$, whereby dissipation increases monotonically with accuracy. 
Thus, while the introduction of explicit sensing preserves the precision-dissipation tradeoff structure put forward in Ref.~\cite{cocconi2025dissipation} (where $\dot{\Sigma}_{\rm opt} - \Pe \simeq \sigma_{r, \rm opt}^{-2}$), it modifies its characteristic asymptotic scaling relation, rendering the low precision regime comparatively more costly.
In addition, in this regime the Pareto front is characterized by the relation $\sigma_{\varphi, \rm opt} \propto \sigma^{1/4}_{r, \rm opt}$,
which quantifies the required improvements in sensor precision as the localization accuracy increases.

Figure~\ref{fig:pareto_point_target}(a) juxtaposes the asymptotic front~\eqref{eq:small_kapt_tradeoff} against the exact numerical minimization of $\dot{\Sigma}-\Pe$ at fixed $\kapt_{\rm eff}$, confirming excellent agreement down to $\sigma_r^2 \approx 20$, corresponding to values of $\kapt_{\rm eff} \lesssim 0.5$.
A more general, albeit looser, bound can be constructed directly in terms of the bare alignment strength $\kapt$. Since the localisation error decreases monotonically with improving sensor precision, it is bounded below by its zero sensory error limit, $\sigma^2_{r,\rm opt}(\kapt) \geq \lim_{\sigma_\varphi^2 \to 0} \sigma^2_{r}(\kapt,\sigma^2_\varphi) = \bar\ell^2\Gamma(4+\bar\eta)/\Gamma(2+\bar\eta)$, where $\bar\ell = \ell(\kapt_{\rm eff}=\kapt)$ and $\bar\eta=\eta(\kapt_{\rm eff}=\kapt)$. Conversely, the dissipation is bounded below by its infinite sensory error limit, $\dot\Sigma_{\rm opt}(\kapt) \geq \Pe + \lim_{\sigma_\varphi^2 \to \infty} \dot\Sigma(\kapt,\sigma_\varphi^2) = \Pe + \kapt^2/2$. 

We further illustrate in Fig.~\ref{fig:pareto_point_target}(b) the energetic bookkeeping that underlies the Pareto front.
At optimality, the dissipation associated with actuation and sensing becomes comparable, indicating a non-trivial allocation of thermodynamic resources across distinct functionalities.
For high accuracy, where the particle remains close to the target,
$\dot{\Sigma}_{\rm act, pinned}$ and $\dot{\Sigma}_{\rm act, \nabla}$ are also comparable, highlighting the presence of strong policy gradients.
On the other hand, $\dot{\Sigma}_{\rm act, \nabla} \to 0$ for large $\sigma_r^2$ values, reflecting the fact that the particle spends more time away from the target where $\theta^*(r)$ is more uniform.

Validating the results of the multiscale expansion by direct numerical simulations of the Langevin dynamics~\eqref{eq:governing_langevin} actually proves challenging in the point target setup.
This is due to the slow convergence of the empirical probability density $\rho(r)$ to its $r\to 0$ power-law asymptote---exacerbated by the smallness of the anomalous exponent $\eta$  in the weak-steering regime, where $\eta = 5 \kapt^2_{\rm eff}/32 + \mathcal{O}(\kapt^3_{\rm eff})$ ---which demands unfeasibly long simulations of the full dynamics. 
In the following section, we thus introduce a regularised point-target geometry amenable to direct computational checks.

\subsection{Regularised point target: localisation within a disc}
\label{sec:target_disc}

We modulate the sensory alignment term, $\mut\sin(\varphi-\theta^*)$ in Eq.~\eqref{eq:curr_varphi}, with a position-dependent factor $\gamma(r)$ that vanishes as the particle enters the target. 
Physically, such modulation could model an extended domain near $\bm{r}_{\rm{target}}$ with safe or nutrient-rich conditions, such that sensing and navigation are no more necessary.
In what follows, we use the hard cutoff $\gamma(r)=\Theta(r-R)$, where $\Theta$ is the Heaviside function.

The stationary density is obtained by solving the marginal Fokker-Planck equation~\eqref{eq:pdf_rho_target_gen} with a space-dependent effective alignment strength $\kapt_{\rm eff}(r) = \kapt c_0(\sigma^{-2}_\varphi(r))$ which vanishes for $r<R$. This is because the dimensionless sensory error $\sigma^2_\varphi(r) = D_\varphi/[D_\theta \mut \gamma(r)]$ now diverges for $r < R$, while $c_0(0) = 0$.
Solving Eq.~\eqref{eq:pdf_rho_target_gen} piecewise while imposing
continuity at $r=R$ and normalization yields
\begin{equation}
    \rho_{\rm reg}(r) = \frac{2 r }{R^2 \left[ 1 + 2 e^{R/\ell} \left( \frac{\ell}{R}\right)^{2+\eta} \Gamma\left( 2+\eta, \frac{R}{\ell}\right) \right]}
    \begin{cases}
        1 &\text{ if }\quad  0 \leq r \leq R\\
        \left( \frac{r}{R}\right)^\eta e^{-(r-R)/\ell} &\text{otherwise}\\
    \end{cases}~, \label{eq:rho_part_reg}
\end{equation}
where $\Gamma\left( \alpha, x\right)$ is the upper incomplete Gamma function.
Accordingly, the stationary probability for the particle to be inside the target disc is
\begin{equation}
    \label{eq:pi_in}
    \pi_{\rm in} = \frac{1}{1 + 2 e^{R/\ell}E_{-1-\eta}(R/\ell)}~,
\end{equation}
where $E_n(z) =\int_1^\infty dt e^{-zt} t^{-n}$ is an exponential integral function. 
As detailed in~\ref{app:target}, the sensing dissipation~\eqref{eq:epr_sens_fast} reads
\begin{equation}
    \label{eq:sigma_sens_reg}
    \dot{\Sigma}_{\rm sens, reg} = \frac{c_0(\sigma_\varphi^{-2})}{\sigma_\varphi^2\Pe } \; \frac{2 e^{R/\ell}}{R^2} \; \frac{\left(\frac{\ell}{R}\right)^\eta \Gamma\left( \eta,\frac{R}{\ell} \right)}{1 + 2 e^{R/\ell} \left( \frac{\ell}{R}\right)^{2+\eta} \Gamma \left( 2+\eta, \frac{R}{\ell} \right)} ~,
\end{equation}
which for small $R$ asymptotes to
\begin{equation}
\dot{\Sigma}_{\rm sens, reg} \underset{R \to 0}{\simeq}  
\frac{c_0(\sigma_\varphi^{-2})}{\sigma_\varphi^2\Pe } \, \frac{\Gamma(\eta)}{\ell^2 \Gamma(2+\eta)} \left[
1 - \frac{1}{\eta\,\Gamma(\eta)}\left(\frac{R}{\ell}\right)^{\eta}\right]~.\label{eq:asymp_sens_reg}
\end{equation}
It is clear that taking $R \to 0$ in Eq.~\eqref{eq:asymp_sens_reg} recovers the result of the unregularised case.
Using Eq.~\eqref{eq:rho_part_reg}, we directly evaluate the variance of the particle displacement from the target center: 
\begin{equation}
    \label{eq:variance_reg}
    \sigma_{r,\rm reg}^2 = \frac{R^2\left( 1 + 4 e^{R/\ell} E_{-3-\eta}(R/\ell)\right)}{2 + 4 e^{R/\ell} E_{-1-\eta}(R/\ell)}~,
\end{equation} 
which likewise converges to its unregularised form, Eq.~\eqref{eq:theory}, as $R\to 0$.

To complete the thermodynamic characterisation, we derive the regularised actuation contributions to the total entropy production.
The pinned component, cf.~Eq.~\eqref{eq:epr_act_pinned}, splits into two terms weighted by the probability of finding the particle inside ($\pi_{\rm in}$) or outside ($1-\pi_{\rm in}$) of the disc:
\begin{equation}
    \label{eq:act_pinned_reg}
    \dot{\Sigma}_{\rm act,pinned,reg} = \pi_{\rm in} \frac{\kapt^2}{2} + (1-\pi_{\rm in}) \dot{\Sigma}_{\rm act,pinned},
\end{equation}
since $\lim_{\sigma_\varphi^2 \to \infty} \dot{\Sigma}_{\rm act,pinned} = \kapt^2/2$. 
The $\mathcal{O}(\nabla)$ correction, $\dot{\Sigma}_{{\rm act},\nabla,{\rm reg}}$, generated by the moment expansion is instead obtained upon replacing $\langle \nabla \cdot \hu^* \rangle$ in Eq.~\eqref{eq:epr_act_nabla} with 
\begin{equation}
    \label{eq:div_ustar_reg}
    \langle \Theta(r-R)\nabla\cdot\hu^*\rangle = - \frac{2 e^{R/\ell }E_{-\eta}(R/\ell)}{R(1 + 2 e^{R/\ell} E_{-1-\eta}(R/\ell))}~,
\end{equation} 
which remains finite as $R\to 0$ for any $\eta \geq 0$. Thus, regularisation preserves the qualitative structure of Sec.~\ref{sec:target_point} results.

\begin{figure}[t!]
    \centering
    \includegraphics[width=\linewidth]{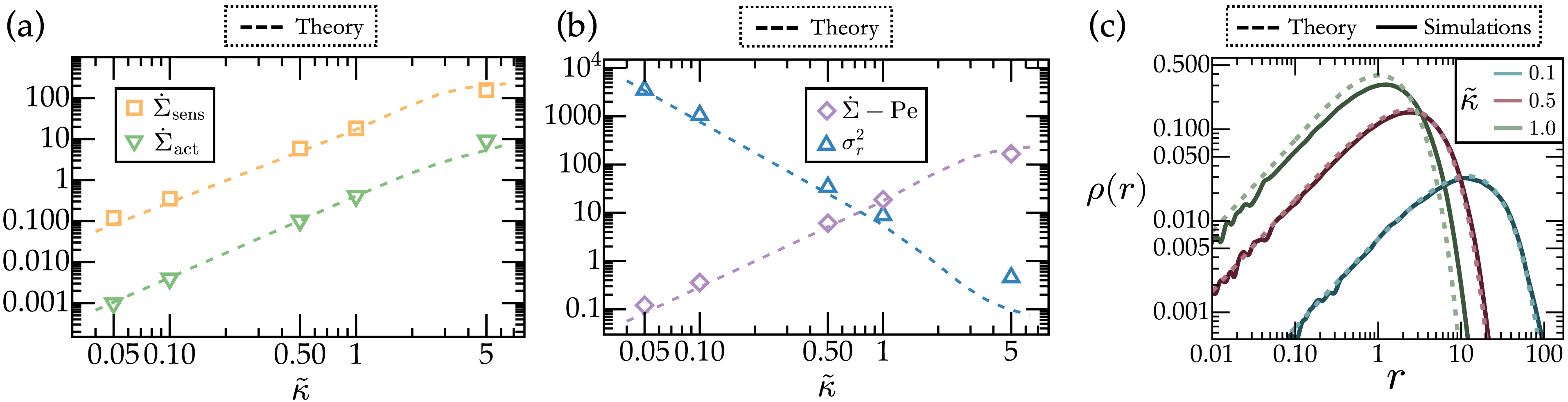}
    \caption{Localisation within a target disc. 
    (a) Thermodynamic costs of actuation, (green), and sensing (orange), as functions of the renormalised alignment strength $\tilde\kappa$.
    Dashed lines indicate the theoretical predictions [cf.~(\ref{eq:sigma_sens_reg}-\ref{eq:div_ustar_reg})] while symbols show values obtained from numerical simulations of Eqs.~\eqref{eq:governing_langevin}. 
    (b) same as (a) for the total entropy production (purple), and variance of the particle displacement from the target center (blue). 
    (c) The stationary density $\rho(r)$ profile obtained from Eq.~\eqref{eq:rho_part_reg} (dashed lines) and Langevin dynamics simulations (solid lines). 
    In all panels, $R=0.1$, $\tilde\mu=50$, $\sigma_\varphi^2=0.01$, and $\Pe=10$.}
    \label{fig:target_disc}
\end{figure}

As anticipated, 
the total entropy production $\dot{\Sigma}$ in the regularised problem depends only weakly on the characteristics of
the density for $r \to 0$, enabling a quantitative comparison between theory and numerics.
Our analytical predictions for the stationary density---Eq.~\eqref{eq:rho_part_reg}---as well as sensing and actuation thermodynamic costs---Eqs.~\eqref{eq:sigma_sens_reg} and~(\ref{eq:act_pinned_reg}, \ref{eq:div_ustar_reg}), respectively---are confronted to data obtained from direct numerical simulations of Eqs.~\eqref{eq:governing_langevin} in Fig.~\ref{fig:target_disc}.
We find excellent agreement in all cases for $\tilde{\kappa}\lesssim 1$, which validates the moment-expansion and fast-sensing approximations in the target geometry.
In this case, where $\kapt$ is varied while keeping the sensor precision $\sigma_\varphi$ low,
the total dissipation is naturally dominated by $\dot{\Sigma}_{\rm sens}$ (Fig.~\ref{fig:target_disc}(a)), at odds with the optimal resource allocation reported in Fig.~\ref{fig:pareto_point_target}(b).
As a consequence, the data shown in Fig.~\ref{fig:target_disc}(b) remain well above the bound defined by the Pareto front in Fig.~\ref{fig:pareto_point_target}(a).

\subsection{Navigation along a target path}
\label{sec:target_path}

We finally consider an agent tasked with maintaining itself near a prescribed 1D target path while navigating along the latter in a preferred direction. 
This scenario proves particularly relevant for stochastic optimal navigation, where energy- or time-minimising strategies often reduce to trajectory tracking~\cite{piro2022optimal,muinos2021reinforcement,piro2024energetic}.

For tractability, we assume that the path's typical radius of curvature is large relative to other characteristic length scales of the motion, permitting a straight-line approximation. We further require the policy $\hu^*$ to be independent of the longitudinal coordinate and ``localising'', namely ${\rm sgn}[\cos\theta^*(y)] = -{\rm sgn}(y)$ where $y$ denotes the signed transverse displacement from the target path. Longitudinal translational invariance then allows integrating the parallel coordinate from the effective steady-state Fokker-Planck equation~\eqref{eq:eff_fp_rho}, yielding a 1D description in $y$:
\begin{align}
    \partial_y \ln \rho(y) 
    &= \frac{c_0(\kapt_{\rm eff})\cos\theta^* + \frac{1}{2}[c_3(\kapt_{\rm eff})-c_4(\kapt_{\rm eff})]\partial_y \cos^2\theta^*}{\Pe^{-1} + c_1(\kapt_{\rm eff}) + c_2(\kapt_{\rm eff}) \cos^2\theta^*}~.
\end{align}
It follows by direct integration that the probability density of the orthogonal displacement is given, for a generic policy $\hu^*$ satisfying the assumptions specified above, by
\begin{equation}
    \rho(y;\hu^*) = \rho_0 \exp\left[ \int^y dy' \frac{c_0(\kapt_{\rm eff})\cos\theta^*(y') + \frac{1}{2}[c_3(\kapt_{\rm eff})-c_4(\kapt_{\rm eff})]\partial_{y'} \cos^2\theta^*(y')}{\Pe^{-1} + c_1(\kapt_{\rm eff}) + c_2(\kapt_{\rm eff}) \cos^2\theta^*(y')}\right] \label{eq:rho_path}
\end{equation}
with normalization $\rho_0$. Thermodynamic quantities follow by evaluating the expectations in Eqs.~\eqref{eq:epr_act_pinned}, \eqref{eq:epr_act_nabla} and \eqref{eq:epr_sens_fast} using \eqref{eq:rho_path}.
In this geometry, a key new metric emerges: the mean longitudinal velocity along the path,
\begin{align}
    V &\equiv \left\langle \hat{\bm{e}}_\parallel \cdot \langle \hu_\theta \rangle_\theta \right\rangle + \mathcal{O}(\nabla^2,\mut^{-1}) \nonumber \\
    &\simeq c_0(\kapt_{\rm eff}) \langle \sin\theta^*\rangle + c_2(\kapt_{\rm eff}) \langle \partial_y(\sin\theta^* \cos\theta^*)\rangle - \frac{c_3(\kapt_{\rm eff})+c_4(\kapt_{\rm eff})}{2} \langle \partial_y \cos^2\theta^*\rangle  \label{eq:mean_v}
\end{align}
using Eq.~\eqref{eq:ave_heading_momexp}. This quantity is related to but not directly derivable from the variance of the transverse displacement $\sigma_y^2$ and quantifies directed progress along the path.

\begin{figure}
    \centering
    \includegraphics[width=\linewidth]{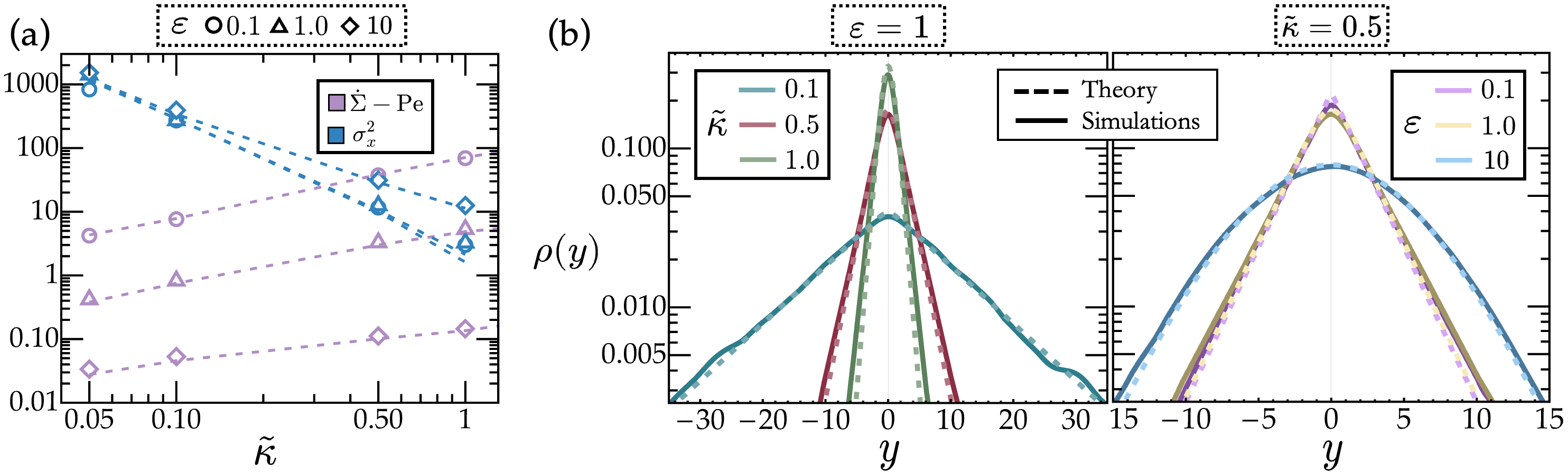}
    \caption{Navigation along a target path. (a) Comparison between theory (dashed curves) and numerical simulations (symbols) for the total entropy production (purple) and of the variance in the transverse direction (blue) are shown as functions of $\kapt$ and for three distinct values of $\varepsilon$. (b) Probability density functions of the orthogonal displacement obtained from the numerical simulations (solid curves) compared to those predicted by Eq.~\eqref{eq:rho_path} (dashed curves) for different values of $\kapt$ (left) and $\varepsilon$ (right). The close match demonstrates the quantitative accuracy of the theoretical framework also in this setup.}
    \label{fig:target_path}
\end{figure}

Unlike the case of a point target, for which $\hu^*$ is generically non differentiable at $r=0$, the present setting admits steering policies that are smooth everywhere. As a consequence, no singular contributions arise in either $\dot\Sigma_{\rm sens}$ or $\dot\Sigma_{\rm act,\nabla}$, and all entropy production terms remain regular.
As a concrete example, we consider the policy $\cos\theta^*(\bm{r})=-\tanh(y/\varepsilon)$ with $\sin\theta^*>0$ (positive longitudinal bias) and confinement length scale $\varepsilon$, reminiscent of the adaptive aligning policy (AAP) introduced in Ref.~\cite{piro2022optimal}.
For this policy, $\nabla\cdot\hu^*=\partial_y \cos\theta^*=-\varepsilon^{-1}{\rm sech}^2(y/\varepsilon)$ and $|\nabla \theta^*|^2 = |\partial_y \theta^*|^2 = \varepsilon^{-2}{\rm sech}^2(y/\varepsilon)$, such that one could in principle calculate the ratio $\dot\Sigma_{\rm sens}/\dot\Sigma_{\rm act,\nabla}$ without knowing $\rho(\bm{r})$. 

Figure~\ref{fig:target_path} compares analytical predictions---obtained by numerically evaluating the integrals defining accuracy and all components of the dissipation with Eq.~\eqref{eq:rho_path}---with direct Langevin simulations of the full dynamics. The agreement is excellent across the explored parameter range, providing further validation of our approximation scheme.

\begin{figure}
    \centering
    \includegraphics[width=\linewidth]{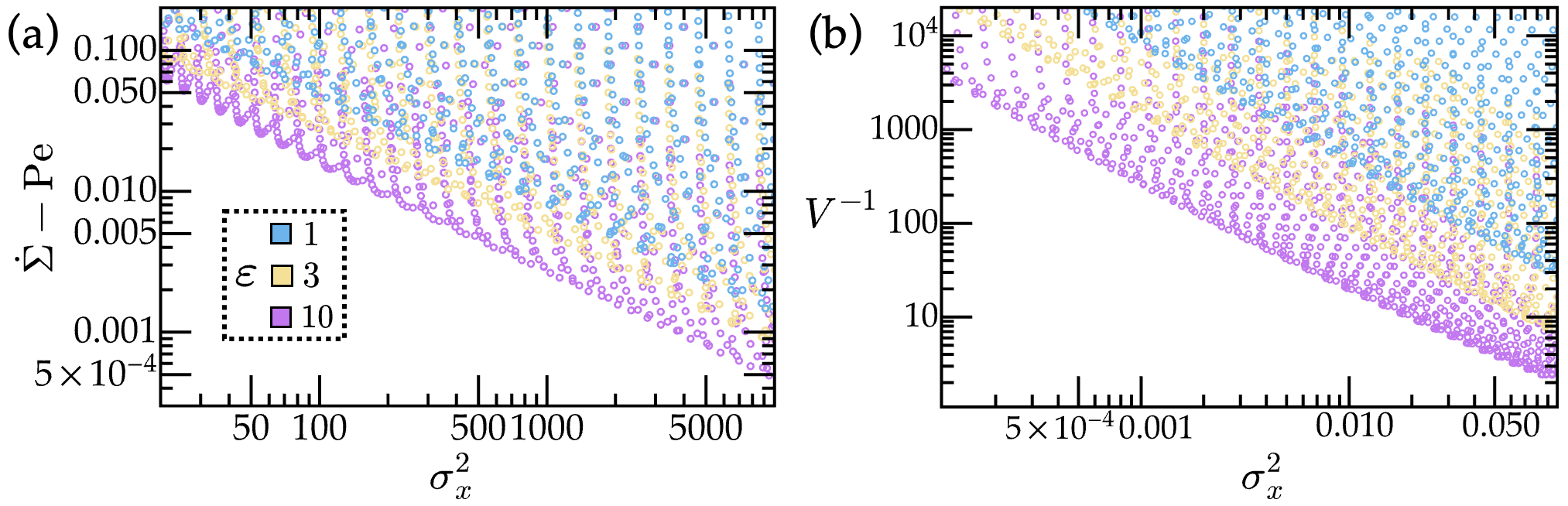}
    \caption{Pareto fronts for navigation along a target path obtained by substituting the density \eqref{eq:rho_path} into the Eqs.~(\ref{eq:epr_act_pinned}--\ref{eq:epr_sens_fast}) and \eqref{eq:mean_v}. The resulting integral expressions are evaluated numerically. To reduce computational costs, we work in the weak steering regime $\kapt \ll 1$ by only retaining corrections of leading order in $\kapt_{\rm eff}$ in the relevant expressions.
    Comparing across different values of the confinement length scale $\varepsilon$, we observe that both the precision-dissipation (a) and speed-dissipation (b) tradeoffs are softened for increasing $\varepsilon$.
    }
    \label{fig:pareto_path}
\end{figure}

We then examine how the Pareto structure identified for point localization is reshaped by the additional geometric parameter $\varepsilon$. 
Figure~\ref{fig:pareto_path} displays $\dot\Sigma-\Pe$ as a function of both the transverse variance $\sigma_y^2$ and the inverse longitudinal speed $V^{-1}$. 
To efficiently scan parameter space, we compute analytical predictions to leading order in small $\kapt_{\rm eff}$. 
We find that increasing $\varepsilon$ systematically ``softens'' the precision/speed–dissipation trade-off,
since a given accuracy can be reached with gradually lower dissipation.
Thus, while the quantitative shape of the frontier depends on $\varepsilon$, its thermodynamic origin and qualitative structure remain unchanged.

\section{Discussion and conclusion}
\label{sec:conclusion}

We introduced a minimal model of a self-steering active particle in which motility, sensing, and actuation are treated on equal thermodynamic footing (Sec.~\ref{sec:model_setup}). By resolving the internal sensor dynamics explicitly and embedding them within a stochastic feedback loop, we were able to decompose the steady-state entropy production into physically interpretable contributions associated with locomotion, control, and sensing (Sec.~\ref{s:epr_framework}). This separation exposes a concrete energetic bookkeeping structure underlying even the simplest form of embodied navigation. By establishing a quantitative bridge between dissipation and multivariate information flow in this spatially extended agent, we also demonstrate that the energetic cost of sensing can be interpreted as a measurable price of information acquisition (Sec.~\ref{s:info_flow}).

Within a controlled hierarchy of timescales, the fast-sensing limit yields an effective drift–diffusion description in which the steering policy acts as a geometric field shaping both transport and fluctuations. This coarse-grained structure allows analytical access to steady-state densities and leads to explicit dissipation–accuracy trade-offs for representative navigation tasks (Sec.~\ref{s:example_policies}). The emergence of Pareto fronts linking energetic expenditure to localisation precision and path-following performance shows that feedback-controlled active motion is constrained by quantitative thermodynamic bounds. Importantly, the persistence of these trade-offs across distinct task geometries indicates that they are not artefacts of a specific policy, but rather reflect generic features of sensorimotor coupling in smart active matter.

Several natural extensions follow. Allowing the steering policy itself to adapt or be learned \cite{muinos2021reinforcement,putzke2023optimal,schneider2019optimal} would introduce additional thermodynamic costs associated with policy acquisition, placing adaptation within the same accounting structure as sensing and actuation. Likewise, taking into account the energetic price of policy negotiation across agents opens a direct route to multi-agent scenarios, where communication \cite{bryant2023physical,yadav2025minimal} and coordination might generate new trade-offs between individual expenditure and collective performance. Such extensions are particularly relevant for the design of autonomous microswimmers \cite{liu2023colloidal,li2017micro} and swarm robotic systems \cite{sun2023mean,dorigo2020reflections}, where sensing, decision-making, and motion must coexist under tight energetic budgets. 

Taken together, our results advance a minimal yet analytically tractable framework for the nonequilibrium energetics of smart active matter~\cite{baulin2025intelligent}. By unifying propulsion, sensing, and feedback control within a single stochastic thermodynamic description, we provide a step toward a quantitative theory in which navigation and decision-making are constrained, organised, and ultimately limited by energy dissipation.

\textbf{Acknowledgements:} We acknowledge support from the Alexander von Humboldt Foundation. This work was supported by the European Research Council (ERC) under the European Union’s Horizon 2020 research and innovation programme (Grant Agreement No. 882340). 

\appendix

\section{Moment expansion}\label{app:mom_exp}

The approximation scheme introduced in Sec.~\ref{sec:approximations} to determine the steady-state solution of the marginal Fokker-Planck equation \eqref{eq:eff_fp_rho} closely follows the steps detailed in Appendix B of Ref.~\cite{cocconi2025dissipation}, which we don't reproduce here for the sake of brevity. It draws on the construction and suitable truncation of factorised solutions of the form
\begin{equation} 
P_{\rm eff}(\bm r,\theta,t) = \rho(\bm r,t) Q(\theta,\bm r) = \rho(\bm r,t) \left[ Q_0(\theta,\bm r) + Q_1(\theta,\bm r) + \ldots \right],
\end{equation}
where $Q_i = {\cal O}(\nabla^i)$ are of increasing order in the spatial gradients. For later use, we introduce the notation $\langle \cdot \rangle_{k} \equiv \int_0^{2\pi}\rmd\theta \, (\cdot) Q_k$ for averages computed using $Q_k$ as a measure. 
Indeed, having integrated out the fast sensory dynamics, the joint probability density $P_{\rm eff}(\bm{r},\theta,t)$ in Eq.~\eqref{eq:p_factorised} of the present work evolves according to the effective Fokker-Planck equation [20, (B.1)] with a renormalised alignment strength $\kapt_{\rm eff} = \kapt c_0(\sigma^{-2}_\varphi)$. Borrowing directly from Ref.~\cite{cocconi2025dissipation}, specifically combining Eqs.~[20, (7)] and [20, (B.13)], we thus have that
\begin{equation}
     \kapt_{\rm eff}^2\langle\sin^2(\theta-\theta^*) \rangle - \kapt_{\rm eff}\langle\cos(\theta-\theta^*) \rangle = - c_0(\kapt_{\rm eff})\langle \nabla \cdot \hu^*\rangle + O(\nabla^2,\mut^{-1})~.
\end{equation}
Note that the combination of trigonometric moments in the left-hand side of the above expression had a specific physical interpretation in Ref.~\cite{cocconi2025dissipation} (as a distinct contribution to the total entropy production), which is not the case here.

Going beyond the results of Ref.~\cite{cocconi2025dissipation}, we now derive the average orientation $\langle \hu_\theta \rangle_\theta$ appearing in Eq.~\eqref{eq:eff_fp_rho} to leading order in the moment expansion but all orders in $\kapt_{\rm eff}$. To do so, we first rewrite $\calB_1(\theta,\bm r)$ from Eq.~[20, (B.6)] as
\begin{equation}
\calB_1(\theta,\bm r) = {\bm F}_\rho(\theta - \theta^*,\bm r)\cdot\nabla\ln\rho_{\rm s} + {\bm F}_\theta(\theta - \theta^*,\bm r)\cdot\nabla\theta^*,
\end{equation}
where
\begin{align*}
{\bm F}_\rho & = Q_0(\theta - \theta^*)\left[ \left(\cos(\theta - \theta^*) - c_0(\kapt) \right)\hu^* + \sin(\theta - \theta^*) \hu^*_\perp \right] , \\
{\bm F}_\theta & = -Q_0'(\theta - \theta^*)\left[ \bm v + \cos(\theta - \theta^*)\hu^* + \sin(\theta - \theta^*)\hu^*_\perp \right] - Q_0(\theta - \theta^*)c_0(\kapt)\hu^*_\perp.
\end{align*}
Hence, using [20, (B.8)], we obtain
\begin{equation}
\calX_1 = \bm X_1^\rho \cdot \nabla\ln\rho_{\rm s} + \bm X_1^\theta \cdot \nabla\theta^*,
\end{equation}
where $\bm X_1^\rho$ and $\bm X_1^\theta$ are calculated by integrating ${\bm F}_\rho$ and ${\bm F}_\theta$, respectively.
To extract the average orientation, we then write $\hu_\theta = \cos(\theta - \theta^*)\hu^* + \sin(\theta - \theta^*)\hu^*_\perp$, such that
\begin{equation}
\langle \hu_\theta \rangle_1 = \left[ \hu^* \bm {c}_\rho + \hu^*_\perp \bm {s}_\rho \right] \cdot \nabla\ln\rho_{\rm s} +
\left[ \hu^* \bm {c}_\theta + \hu^*_\perp \bm {s}_\theta \right] \cdot \nabla\theta^*,
\end{equation}
where $\bm c_{\alpha} = \langle \cos(\theta - \theta^*)  \bm X_1^\alpha(\theta - \theta^*,\bm r) \rangle$
and $\bm s_{\alpha} = \langle \sin(\theta - \theta^*)  \bm X_1^\alpha(\theta - \theta^*,\bm r) \rangle$.
Calculating the averages, we find that the $\bm c$ and $\bm s$ coefficients take the forms
\begin{align*}
\bm c_\rho = c_\rho \hu^*, \qquad \bm c_\theta = c_\theta \hu^*_\perp, \qquad
\bm s_\rho = s_\rho \hu^*_\perp , \qquad \bm s_\theta = s_\theta^v \bm v + s_\theta^* \hu^*.
\end{align*}
Therefore, we can write
\begin{equation}
\langle \hu_\theta \rangle_1 = \left[ s_\rho \bm I + (c_\rho - s_\rho) \hu^* \hu^* \right] \cdot \nabla\ln\rho_{\rm s} +
s_\theta^v (\bm v \cdot \nabla)\hu^* + c_\theta \hu^* (\nabla\cdot\hu^*) - s_\theta^* \hu^*_\perp (\nabla\cdot\hu^*_\perp).
\end{equation}
Recalling that $\langle \hu_\theta \rangle_0 = c_0(\kapt_{\rm eff})\hu^*$ and to first order in gradient, we thus find that the average orientation moment takes the general form
\begin{align}
\langle \hu_\theta \rangle_\theta & = c_0(\kapt_{\rm eff})\hu^* - \left[ c_1(\kapt_{\rm eff})\bm I + c_2(\kapt_{\rm eff})\hu^*\hu^* \right]\cdot\nabla\ln\rho - \kapt c_1(\kapt_{\rm eff}) (\bm v\cdot\nabla)\hu^* \nonumber\\
& + c_3(\kapt_{\rm eff})\hu^*(\nabla\cdot\hu^*) + c_4(\kapt_{\rm eff})\hu^*_\perp(\nabla\cdot\hu^*_\perp) + {\cal O}(\nabla^2,\mut^{-1})~,
\label{eq_u_O1_general}
\end{align}
where $\bm I$ denotes the identity matrix and $\hu^*_\perp = \hu\left(\theta^* + \frac{1}{2}\pi\right)$. The coefficients $c_{i}(\kapt_{\rm eff})$ cannot be expressed in terms of simple functions
but can be evaluated at arbitrary order in $\kapt$.
For example, up to $\mathcal{O}(\kapt_{\rm eff}^{20})$ terms, we have
\begin{small}
\begin{align*}
 c_1(\kapt) & = \frac{1}{2} - \frac{5 \kapt ^2}{32} + \frac{23 \kapt^4}{576} - \frac{677 \kapt ^6}{73728} + \frac{7313 \kapt ^8}{3686400}
- \frac{218491 \kapt^{10}}{530841600} + \frac{863897 \kapt ^{12}}{10404495360} \\
  & - \frac{874088357 \kapt^{14}}{53271016243200} + \frac{27545803997 \kapt ^{16}}{8629904631398400}
- \frac{423385249313 \kapt^{18}}{690392370511872000} + \frac{19488418951523 \kapt ^{20}}{167074953663873024000}, \\
 c_2(\kapt) & = -\frac{\kapt ^2}{8} + \frac{\kapt^4}{16}-\frac{131 \kapt ^6}{6144}+\frac{25 \kapt ^8}{4096}
-\frac{41851 \kapt^{10}}{26542080}+\frac{20209 \kapt ^{12}}{53084160}-\frac{33334307 \kapt^{14}}{380507258880} \\
 & +\frac{133205867 \kapt ^{16}}{6849130659840}-\frac{19173165917 \kapt^{18}}{4566087106560000}
+ \frac{3470122403 \kapt ^{20}}{3913788948480000}, \\
 c_3(\kapt) & = \frac{3 \kapt ^2}{32} -\frac{29 \kapt^4}{576} +\frac{1325 \kapt ^6}{73728} -\frac{19553 \kapt ^8}{3686400}
+\frac{247777 \kapt^{10}}{176947200} -\frac{397169 \kapt ^{12}}{1156055040} +\frac{473737993\kapt ^{14}}{5919001804800}\\
 & -\frac{154836151097 \kapt ^{16}}{8629904631398400} +\frac{13477269234097 \kapt^{18}}{3451961852559360000}
-\frac{693144302217667 \kapt ^{20}}{835374768319365120000}, \\
 c_4(\kapt) & = \frac{3 \kapt ^2}{32} -\frac{17 \kapt^4}{576} +\frac{545 \kapt ^6}{73728} -\frac{6173 \kapt ^8}{3686400}
+\frac{190111 \kapt^{10}}{530841600} -\frac{767717 \kapt ^{12}}{10404495360} +\frac{788938397 \kapt^{14}}{53271016243200}\\
 & -\frac{25160566037 \kapt ^{16}}{8629904631398400} +\frac{390389770937 \kapt^{18}}{690392370511872000}
-\frac{18107708487467 \kapt ^{20}}{167074953663873024000}.
 \end{align*}
\end{small}

\section{Derivation of the dissipation for the regularized target}\label{app:target}

Here, we derive the expression of the dissipation associated with sensing in the care of the extended target.
We thus consider the Langevin dynamics~\eqref{eq:governing_langevin} with $\mut \to \gamma(r) \mut$, where $\gamma(r)=\Theta(r-R)$ is a step function $\gamma(r) = \Theta(r-R)$ over a disc of radius $R$ centered at the target.

A suitably modified version of Eq.~\eqref{eq:epr_sens_alt}, followed by some manipulation of integrals analogous to the unregularised case, yields
\begin{align}
    \dot\Sigma_{\rm sens,reg} = \frac{c_0(\sigma_\varphi^{-2})}{\sigma_\varphi^2} \Big[ &c_0(\kapt_{\rm eff}) \langle \gamma'(r)\rangle + c_3(\kapt_{\rm eff}) \langle \gamma'(r) \nabla \cdot \hu^* \rangle \nonumber \\
    &+ \Pe^{-1} \left( \langle \gamma(r)|\nabla\theta^*|^2 \rangle - \langle \gamma''(r)\rangle - 
    \left\langle \frac{\gamma'(r)}{r}\right\rangle\right)\Big] + \mathcal{O}(\nabla^2) ~.
    \label{eq:regularised_smooth}
\end{align}
Using that $\gamma'(r)=\delta(r-R)$, and integrating by parts reduces the terms in square brackets in the above expression to 
\begin{equation}
    [...] = c_0(\kapt_{\rm eff})  \rho(R) - c_3(\kapt_{\rm eff}) 
    \frac{\rho(R)}{R}+ \Pe^{-1} \left( \langle \Theta(r-R)|\nabla\theta^*|^2\rangle + \rho'(R) - \frac{\rho(R)}{R} \right)~.
\end{equation}
This expression separates into a regularised bulk term $\propto \langle \Theta(r-R)|\nabla\theta^*|^2\rangle$---excluding potentially singular contributions to $\dot\Sigma_{\rm sens}$ from a neighborhood around $r=0$---and boundary corrections $\propto \rho(R)$ or $\rho'(R)$. The latter prove subleading (for any $\eta$) as $R\to 0$, justifying their neglect for small $R$ where the bulk term dominates while numerical convergence is more easily attained.
Setting them to zero and evaluating the bulk contribution from the regularized density profile~\eqref{eq:rho_part_reg}, we recover Eq.~\eqref{eq:sigma_sens_reg} of the main text.
Equations~\eqref{eq:act_pinned_reg} and~\eqref{eq:div_ustar_reg} are finally recovered by following a similar approach.

\section*{Bibliography}

\bibliographystyle{iopart-num}
\bibliography{bibliography}

\providecommand{\newblock}{}
\begin{thebibliography}{10}
\expandafter\ifx\csname url\endcsname\relax
  \def\url#1{{\tt #1}}\fi
\expandafter\ifx\csname urlprefix\endcsname\relax\def\urlprefix{URL }\fi
\providecommand{\eprint}[2][]{\url{#2}}

\bibitem{liu2023colloidal}
Liu A~T, Hempel M, Yang J~F, Brooks A~M, Pervan A, Koman V~B, Zhang G, Kozawa
  D, Yang S, Goldman D~I {\em et~al.\/} 2023 {\em Nature materials\/} {\bf 22}
  1453--1462

\bibitem{li2017micro}
Li J, Esteban-Fern{\'a}ndez~de {\'A}vila B, Gao W, Zhang L and Wang J 2017 {\em
  Science robotics\/} {\bf 2} eaam6431

\bibitem{sun2023mean}
Sun G, Zhou R, Ma Z, Li Y, Gro{\ss} R, Chen Z and Zhao S 2023 {\em Nature
  Communications\/} {\bf 14} 3476

\bibitem{dorigo2020reflections}
Dorigo M, Theraulaz G and Trianni V 2020 {\em Science robotics\/} {\bf 5}
  eabe4385

\bibitem{levine2023physics}
Levine H and Goldman D~I 2023 {\em Soft Matter\/} {\bf 19} 4204--4207

\bibitem{goldman2024robot}
Goldman D~I and Zeb~Rocklin D 2024 {\em Science robotics\/} {\bf 9} eadn6035

\bibitem{stark2021artificial}
Stark H 2021 {\em Science Robotics\/} {\bf 6} eabh1977

\bibitem{keller1971model}
Keller E~F and Segel L~A 1971 {\em Journal of theoretical biology\/} {\bf 30}
  225--234

\bibitem{endres2008accuracy}
Endres R~G and Wingreen N~S 2008 {\em Proceedings of the National Academy of
  Sciences\/} {\bf 105} 15749--15754

\bibitem{daniels2004quorum}
Daniels R, Vanderleyden J and Michiels J 2004 {\em FEMS microbiology reviews\/}
  {\bf 28} 261--289

\bibitem{PapenfortQS}
Papenfort K and Bassler B~L 2016 {\em Nature Reviews Microbiology\/} {\bf 14}
  576--588

\bibitem{ju2025technology}
Ju X, Chen C, Oral C~M, Sevim S, Golestanian R, Sun M, Bouzari N, Lin X, Urso
  M, Nam J~S {\em et~al.\/} 2025 {\em ACS nano\/}

\bibitem{chen2025roadmap}
Chen S, Fan D~E, Fischer P, Ghosh A, G{\"o}pfrich K, Golestanian R, Hess H, Ma
  X, Nelson B~J, Pati{\~n}o~Padial T {\em et~al.\/} 2025 {\em Nature
  nanotechnology\/} {\bf 20} 990--1000

\bibitem{miskin2020electronically}
Miskin M~Z, Cortese A~J, Dorsey K, Esposito E~P, Reynolds M~F, Liu Q, Cao M,
  Muller D~A, McEuen P~L and Cohen I 2020 {\em Nature\/} {\bf 584} 557--561

\bibitem{bo2015thermodynamic}
Bo S, Del~Giudice M and Celani A 2015 {\em Journal of Statistical Mechanics:
  Theory and Experiment\/} {\bf 2015} P01014

\bibitem{lan2012energy}
Lan G, Sartori P, Neumann S, Sourjik V and Tu Y 2012 {\em Nature physics\/}
  {\bf 8} 422--428

\bibitem{sartori2015thermodynamics}
Sartori P and Pigolotti S 2015 {\em Physical Review X\/} {\bf 5} 041039

\bibitem{yu2022energy}
Yu Q, Kolomeisky A~B and Igoshin O~A 2022 {\em Journal of The Royal Society
  Interface\/} {\bf 19} 20210883

\bibitem{yang2021physical}
Yang X, Heinemann M, Howard J, Huber G, Iyer-Biswas S, Le~Treut G, Lynch M,
  Montooth K~L, Needleman D~J, Pigolotti S {\em et~al.\/} 2021 {\em Proceedings
  of the National Academy of Sciences\/} {\bf 118} e2026786118

\bibitem{cocconi2025dissipation}
Cocconi L, Mahault B and Piro L 2025 {\em New Journal of Physics\/} {\bf 27}
  013002

\bibitem{welker2026accuracy}
Welker T and Pietzonka P 2026 {\em arXiv preprint arXiv:2602.13173\/}

\bibitem{olsen2026information}
Olsen K~S, Tarama M and L{\"o}wen H 2026 {\em arXiv preprint
  arXiv:2602.23988\/}

\bibitem{muinos2021reinforcement}
Muinos-Landin S, Fischer A, Holubec V and Cichos F 2021 {\em Science
  Robotics\/} {\bf 6} eabd9285

\bibitem{putzke2023optimal}
Putzke M and Stark H 2023 {\em The European Physical Journal E\/} {\bf 46} 48

\bibitem{seifert2012stochastic}
Seifert U 2012 {\em Reports on progress in physics\/} {\bf 75} 126001

\bibitem{solon2015active}
Solon A~P, Cates M~E and Tailleur J 2015 {\em The European Physical Journal
  Special Topics\/} {\bf 224} 1231--1262

\bibitem{goh2022noisy}
Goh S, Winkler R~G and Gompper G 2022 {\em New Journal of Physics\/} {\bf 24}
  093039

\bibitem{de2013non}
De~Groot S~R and Mazur P 2013 {\em Non-equilibrium thermodynamics\/} (Courier
  Corporation)

\bibitem{chen2025numerical}
Chen L and Pruessner G 2025 {\em Physical Review Research\/} {\bf 7} 033295

\bibitem{cocconi2020entropy}
Cocconi L, Garcia-Millan R, Zhen Z, Buturca B and Pruessner G 2020 {\em
  Entropy\/} {\bf 22} 1252

\bibitem{pavliotis2008multiscale}
Pavliotis G and Stuart A 2008 {\em Multiscale methods: averaging and
  homogenization\/} (Springer Science \& Business Media)

\bibitem{sartori2014thermodynamic}
Sartori P, Granger L, Lee C~F and Horowitz J~M 2014 {\em PLoS computational
  biology\/} {\bf 10} e1003974

\bibitem{allahverdyan2009thermodynamic}
Allahverdyan A~E, Janzing D and Mahler G 2009 {\em Journal of Statistical
  Mechanics: Theory and Experiment\/} {\bf 2009} P09011

\bibitem{loos2020irreversibility}
Loos S~A and Klapp S~H 2020 {\em New Journal of Physics\/} {\bf 22} 123051

\bibitem{srinivasa2005review}
Srinivasa S 2005 {\em Univ. of Notre Dame, Notre Dame, Indiana\/} {\bf 2}

\bibitem{kullback1951information}
Kullback S and Leibler R~A 1951 {\em The annals of mathematical statistics\/}
  {\bf 22} 79--86

\bibitem{lozano2016phototaxis}
Lozano C, Ten~Hagen B, L{\"o}wen H and Bechinger C 2016 {\em Nature
  communications\/} {\bf 7} 12828

\bibitem{frangipane2018dynamic}
Frangipane G, Dell'Arciprete D, Petracchini S, Maggi C, Saglimbeni F, Bianchi
  S, Vizsnyiczai G, Bernardini M~L and Di~Leonardo R 2018 {\em Elife\/} {\bf 7}
  e36608

\bibitem{piro2022optimal}
Piro L, Mahault B and Golestanian R 2022 {\em New Journal of Physics\/} {\bf
  24} 093037

\bibitem{piro2024energetic}
Piro L, Vilfan A, Golestanian R and Mahault B 2024 {\em Physical Review
  Research\/} {\bf 6} 013274

\bibitem{schneider2019optimal}
Schneider E and Stark H 2019 {\em Europhysics Letters\/} {\bf 127} 64003

\bibitem{bryant2023physical}
Bryant S~J and Machta B~B 2023 {\em Physical review letters\/} {\bf 131} 068401

\bibitem{yadav2025minimal}
Yadav A and Wolpert D 2025 {\em Physical Review Research\/} {\bf 7} 043324

\bibitem{baulin2025intelligent}
Baulin V~A, Giacometti A, Fedosov D~A, Ebbens S, Varela-Rosales N~R, Feliu N,
  Chowdhury M, Hu M, F{\"u}chslin R, Dijkstra M {\em et~al.\/} 2025 {\em Soft
  Matter\/} {\bf 21} 4129--4145

\end{thebibliography}

\end{document}